\documentclass[10pt,a4paper,onecolumn]{article}
\usepackage{marginnote}
\usepackage{graphicx}
\usepackage{xcolor}
\usepackage{authblk,etoolbox}
\usepackage{titlesec}
\usepackage{calc}
\usepackage{tikz}
\usepackage{caption}
\usepackage{tcolorbox}
\usepackage{amssymb,amsmath}
\usepackage{ifxetex,ifluatex}
\usepackage{seqsplit}
\usepackage{xstring}
\usepackage[pdfa]{hyperref}
\usepackage{hyperxmp}
\hypersetup{unicode=true,
            pdfapart=3,
            pdfaconformance=B,
            pdftitle={Cthulhu: An Open Source Molecular and Atomic Cross Section Computation Code for Substellar Atmospheres},
            pdfauthor={Arnav Agrawal, Ryan J. MacDonald},
            pdfpublication={Journal of Open Source Software},
            pdfpublisher={Open Journals},
            pdfissn={2475-9066},
            pdfpubtype={journal},
            pdfvolumenum={9},
            pdfissuenum={102},
            pdfdoi={10.21105/joss.06894},
            pdfcopyright={Copyright (c) 2024, Arnav Agrawal, Ryan J. MacDonald},
            pdflicenseurl={http://creativecommons.org/licenses/by/4.0/},
            colorlinks=true,
            linkcolor=[rgb]{0.0, 0.5, 1.0},
            citecolor=Blue,
            urlcolor=[rgb]{0.0, 0.5, 1.0},
            breaklinks=true
}

\usepackage{float}
\let\origfigure\figure
\let\endorigfigure\endfigure
\renewenvironment{figure}[1][2] {
    \expandafter\origfigure\expandafter[H]
} {
    \endorigfigure
}

\usepackage{orcidlink}
\usepackage{fixltx2e} 
\usepackage[
  backend=biber,
]{biblatex}
\bibliography{Cthulhu}{}

\let\textttOrig=\texttt
\def\texttt#1{\expandafter\textttOrig{\seqsplit{#1}}}
\renewcommand{\seqinsert}{\ifmmode
  \allowbreak
  \else\penalty6000\hspace{0pt plus 0.02em}\fi}

\makeatletter
\let\href@Orig=\href
\def\href@Urllike#1#2{\href@Orig{#1}{\begingroup
    \def\Url@String{#2}\Url@FormatString
    \endgroup}}
\def\href@Notdoi#1#2{\def\tempa{#1}\def\tempb{#2}%
  \ifx\tempa\tempb\relax\href@Urllike{#1}{#2}\else
  \href@Orig{#1}{#2}\fi}
\def\href#1#2{%
  \IfBeginWith{#1}{https://doi.org}%
  {\href@Urllike{#1}{#2}}{\href@Notdoi{#1}{#2}}}
\makeatother

\newlength{\cslhangindent}
\newlength{\csllabelwidth}
\newenvironment{CSLReferences}[3] 
 {
  \setlength{\parindent}{0pt}
  \ifodd #1 \everypar{\setlength{\hangindent}{\cslhangindent}}\ignorespaces\fi
  \ifnum #2 > 0
  \setlength{\parskip}{#2\baselineskip}
  \fi
 }%
 {}
\usepackage{calc}

\usepackage[top=3.5cm, bottom=3cm, right=1.5cm, left=1.0cm,
            headheight=2.2cm, reversemp, includemp, marginparwidth=4.5cm]{geometry}

\titleformat{\section}
  {\normalfont\sffamily\Large\bfseries}
  {}{0pt}{}
\titleformat{\subsection}
  {\normalfont\sffamily\large\bfseries}
  {}{0pt}{}
\titleformat{\subsubsection}
  {\normalfont\sffamily\bfseries}
  {}{0pt}{}
\titleformat*{\paragraph}
  {\sffamily\normalsize}

\usepackage{fancyhdr}
\makeatletter
\let\ps@plain\ps@fancy
\fancyheadoffset[L]{4.5cm}
\fancyfootoffset[L]{4.5cm}

\definecolor{linky}{rgb}{0.0, 0.5, 1.0}

\newtcolorbox{repobox}
   {colback=red, colframe=red!75!black,
     boxrule=0.5pt, arc=2pt, left=6pt, right=6pt, top=3pt, bottom=3pt}

\newcommand{\ExternalLink}{%
   \tikz[x=1.2ex, y=1.2ex, baseline=-0.05ex]{%
       \begin{scope}[x=1ex, y=1ex]
           \clip (-0.1,-0.1)
               --++ (-0, 1.2)
               --++ (0.6, 0)
               --++ (0, -0.6)
               --++ (0.6, 0)
               --++ (0, -1);
           \path[draw,
               line width = 0.5,
               rounded corners=0.5]
               (0,0) rectangle (1,1);
       \end{scope}
       \path[draw, line width = 0.5] (0.5, 0.5)
           -- (1, 1);
       \path[draw, line width = 0.5] (0.6, 1)
           -- (1, 1) -- (1, 0.6);
       }
   }

\patchcmd{\@maketitle}{center}{flushleft}{}{}
\patchcmd{\@maketitle}{center}{flushleft}{}{}
\patchcmd{\@maketitle}{\LARGE}{\LARGE\sffamily}{}{}
\def\maketitle{{%
  
  \AB@maketitle}}
\makeatletter
\renewcommand\AB@affilsepx{ \protect\Affilfont}
\renewcommand\AB@affilnote[1]{{\bfseries #1}\hspace{3pt}}
\renewcommand{\affil}[2][]%
   {\newaffiltrue\let\AB@blk@and\AB@pand
      \if\relax#1\relax\def\AB@note{\AB@thenote}\else\def\AB@note{#1}%
        \setcounter{Maxaffil}{0}\fi
        \begingroup
        \let\href=\href@Orig
        \let\texttt=\textttOrig
        \let\protect\@unexpandable@protect
        \def\thanks{\protect\thanks}\def\footnote{\protect\footnote}%
        \@temptokena=\expandafter{\AB@authors}%
        {\def\\{\protect\\\protect\Affilfont}\xdef\AB@temp{#2}}%
         \xdef\AB@authors{\the\@temptokena\AB@las\AB@au@str
         \protect\\[\affilsep]\protect\Affilfont\AB@temp}%
         \gdef\AB@las{}\gdef\AB@au@str{}%
        {\def\\{, \ignorespaces}\xdef\AB@temp{#2}}%
        \@temptokena=\expandafter{\AB@affillist}%
        \xdef\AB@affillist{\the\@temptokena \AB@affilsep
          \AB@affilnote{\AB@note}\protect\Affilfont\AB@temp}%
      \endgroup
       \let\AB@affilsep\AB@affilsepx
}
\makeatother

\renewcommand\Affilfont{\sffamily\small\mdseries}
\ifnum 0\ifxetex 1\fi\ifluatex 1\fi=0 
  \usepackage[T1]{fontenc}
  \usepackage[utf8]{inputenc}

\else 
  \ifxetex
    \usepackage{mathspec}
    \usepackage{fontspec}

  \else
    \usepackage{fontspec}
  \fi
  \defaultfontfeatures{Ligatures=TeX,Scale=MatchLowercase}

\fi
\IfFileExists{upquote.sty}{\usepackage{upquote}}{}
\IfFileExists{microtype.sty}{%
\usepackage{microtype}
\UseMicrotypeSet[protrusion]{basicmath} 
}{}

\let\addcontentslineOrig=\addcontentsline
\def\addcontentsline#1#2#3{\bgroup
  \let\texttt=\textttOrig\addcontentslineOrig{#1}{#2}{#3}\egroup}
\let\markbothOrig\markboth
\def\markboth#1#2{\bgroup
  \let\texttt=\textttOrig\markbothOrig{#1}{#2}\egroup}
\let\markrightOrig\markright
\def\markright#1{\bgroup
  \let\texttt=\textttOrig\markrightOrig{#1}\egroup}

\usepackage{graphicx,grffile}
\makeatletter
\def\maxwidth{\ifdim\Gin@nat@width>\linewidth\linewidth\else\Gin@nat@width\fi}
\def\maxheight{\ifdim\Gin@nat@height>\textheight\textheight\else\Gin@nat@height\fi}
\makeatother
\setkeys{Gin}{width=\maxwidth,height=\maxheight,keepaspectratio}
\IfFileExists{parskip.sty}{%
\usepackage{parskip}
}{
\setlength{\parindent}{0pt}
\setlength{\parskip}{6pt plus 2pt minus 1pt}
}
\ifx\paragraph\undefined\else
\let\oldparagraph\paragraph
\renewcommand{\paragraph}[1]{\oldparagraph{#1}\mbox{}}
\fi
\ifx\subparagraph\undefined\else
\let\oldsubparagraph\subparagraph
\renewcommand{\subparagraph}[1]{\oldsubparagraph{#1}\mbox{}}
\fi

\title{Cthulhu: An Open Source Molecular and Atomic Cross Section
Computation Code for Substellar Atmospheres}

        \author[1, 2]{Arnav Agrawal\,\orcidlink{0000-0003-2944-0600}\,}
          \author[3, 4, 2]{Ryan J. MacDonald\,\orcidlink{0000-0003-4816-3469}\,}
    
      \affil[1]{Johns Hopkins University Applied Physics Laboratory,
11100 Johns Hopkins Rd., Laurel, MD 20723, USA}
      \affil[2]{Department of Astronomy and Carl Sagan Institute,
Cornell University, 122 Sciences Drive, Ithaca, NY 14853, USA}
      \affil[3]{Department of Astronomy, University of Michigan, 1085 S.
University Ave., Ann Arbor, MI 48109, USA}
      \affil[4]{NHFP Sagan Fellow}
  \date{\vspace{-7ex}}

\begin{document}
\maketitle

\marginpar{

  \begin{flushleft}
  \sffamily\small

  {\bfseries DOI:} \href{https://doi.org/10.21105/joss.06894}{\color{linky}{10.21105/joss.06894}}

  \vspace{2mm}

  {\bfseries Software}
  \begin{itemize}
    \setlength\itemsep{0em}
    \item \href{https://github.com/openjournals/joss-reviews/issues/6894}{\color{linky}{Review}} \ExternalLink
    \item \href{https://github.com/MartianColonist/Cthulhu}{\color{linky}{Repository}} \ExternalLink
    \item \href{https://doi.org/10.5281/zenodo.13888354}{\color{linky}{Archive}} \ExternalLink
  \end{itemize}

  \vspace{2mm}

  \par\noindent\hrulefill\par

  \vspace{2mm}

  {\bfseries Editor:} \href{https://dfm.io/}{Dan Foreman-Mackey}\ExternalLink\orcidlink{0000-0002-9328-5652} \\
  \vspace{1mm}
    {\bfseries Reviewers:}
  \begin{itemize}
  \setlength\itemsep{0em}
    \item \href{https://github.com/LorenzoMugnai}{@LorenzoMugnai}
    \item \href{https://github.com/arjunsavel}{@arjunsavel}
    \end{itemize}
    \vspace{2mm}

  {\bfseries Submitted:} 23 February 2024\\
  {\bfseries Published:} 05 October 2024

  \vspace{2mm}
  {\bfseries License}\\
  Authors of papers retain copyright and release the work under a Creative Commons Attribution 4.0 International License (\href{http://creativecommons.org/licenses/by/4.0/}{\color{linky}{CC BY 4.0}}).

  \end{flushleft}
}

\hypertarget{summary}{%
\section{Summary}\label{summary}}

Atmospheric studies of exoplanets and brown dwarfs are a cutting-edge and rapidly evolving area of astrophysics research. Powerful new telescopes, such as the James Webb Space Telescope (JWST, \protect\hyperlink{Gardner:2023}{Gardner et al., 2023}) and the upcoming Extremely Large Telescopes (\protect\hyperlink{Skidmore:2015}{Skidmore et al., 2015}; \protect\hyperlink{Fanson:2022}{Fanson et al., 2022}; \protect\hyperlink{Padovani:2023}{Padovani \& Cirasuolo, 2023}), are able to capture in detail spectra of planets and brown dwarfs and thereby probe their chemical composition and physical properties. Calculating models of exoplanet or brown dwarf spectra requires knowledge of the
wavelength-dependent absorption of light (cross sections) by the molecules and atoms in the atmosphere (\protect\hyperlink{Seager:2010}{Seager, 2010}; \protect\hyperlink{Heng:2015}{Heng, 2017}). Reliably calculating spectra of substellar atmospheres requires accurate cross sections, without which measurements of chemical abundances and other atmospheric properties can be biased (e.g., \protect\hyperlink{Hedges:2016}{Hedges \& Madhusudhan, 2016}; \protect\hyperlink{Gharib-Nezhad:2019}{Gharib-Nezhad \& Line, 2019}; \protect\hyperlink{Anisman:2022}{Anisman et al., 2022}).

Cross sections are typically pre-computed on a grid of pressures and temperatures from large databases of quantum mechanical transitions (line lists), such as ExoMol (\protect\hyperlink{Tennyson:2020}{Tennyson et al., 2020}), HITRAN (\protect\hyperlink{Gordon:2022}{Gordon et al., 2022}), HITEMP (\protect\hyperlink{Rothman:2010}{Rothman et al., 2010}), and VALD (\protect\hyperlink{Pakhomov:2017}{Pakhomov et al., 2017}). However, calculating cross sections from line lists is often computationally demanding and has required complex and specialised tools. We aim here to lower the access barrier for users to learn how to calculate molecular and atomic cross sections.

\texttt{Cthulhu} is a pure Python package that rapidly calculates cross sections from atomic and molecular line lists. \texttt{Cthulhu} includes modules to automatically download molecular line lists from online databases and compute cross sections on a user-specified temperature, pressure, and wavenumber grid. \texttt{Cthulhu} requires only CPUs and can run on a user's laptop (for smaller line lists with \textless{} 100 million lines) or on a large cluster in parallel (for many billion lines). \texttt{Cthulhu} includes in-depth Jupyter tutorials in the online documentation. Finally, \texttt{Cthulhu} is intended not only for research purposes, but as an educational tool to demystify the process of making cross sections for atmospheric models.

\hypertarget{statement-of-need}{%
\section{Statement of Need}\label{statement-of-need}}

JWST has recently significantly expanded the number of exoplanet and brown dwarfs with high-quality spectra spanning a wide wavelength range (e.g., \protect\hyperlink{Beiler:2023}{Beiler et al., 2023}; \protect\hyperlink{Miles:2023}{Miles et al., 2023}; \protect\hyperlink{Carter:2024}{Carter et al., 2024}; \protect\hyperlink{Welbanks:2024}{Welbanks et al., 2024}). High-fidelity spectra motivate detailed intercomparisons of exoplanet and brown dwarf modelling codes (e.g. \protect\hyperlink{Barstow:2020}{Barstow et al., 2020}), which often identify opacity database differences as a key modelling limitation. Ground-based high spectral resolution datasets (e.g., \protect\hyperlink{Snellen:2010}{Snellen et al., 2010}; \protect\hyperlink{Birkby:2017}{Birkby et al., 2017}; \protect\hyperlink{Pelletier:2023}{Pelletier et al., 2023}) also critically rely on up-to-date opacity data, since older inaccurate line lists can lead to non-detections of molecules via cross-correlation (e.g. \protect\hyperlink{Merritt:2020}{Merritt et al., 2020}) or propagate into retrieved chemical abundances (e.g. \protect\hyperlink{Brogi:2019}{Brogi \& Line, 2019}). However, despite the key need to continually refine exoplanet and brown dwarf models with the latest state-of-the-art opacities, the process of calculating molecular and atomic cross sections is a non-trivial task that is typically outside the speciality of many exoplanet and brown dwarf researchers.

We have built \texttt{Cthulhu} to provide a user-friendly tool for beginners to learn how to work with the most commonly used line list databases and to readily calculate molecular and atomic cross sections.
There are other open source codes that can calculate cross sections, such as HELIOS-K (\protect\hyperlink{Grimm:2015}{Grimm \& Heng, 2015}; \protect\hyperlink{Grimm:2021}{Grimm et al., 2021}) and ExoCross (\protect\hyperlink{Yurchenko:2018}{Yurchenko et al., 2018}), both of which offer impressive computational performance and are excellent tools for experts to calculate cross sections. However, HELIOS-K requires Nvidia GPUs to run while ExoCross is built in Fortran, both of which can pose accessibility issues for beginners. We offer \texttt{Cthulhu}, a pure Python code designed to run
on CPUs, as a user-friendly entry point into the world of cross sections for substellar atmospheres.

\hypertarget{computing-molecular-and-atomic-cross-sections-with-cthulhu}{%
\section{\texorpdfstring{Computing Molecular and Atomic Cross Sections
with Cthulhu}{Computing Molecular and Atomic Cross Sections with Cthulhu}}\label{computing-molecular-and-atomic-cross-sections-with-cthulhu}}

\begin{figure}[t!]
\centering
\vspace{-0.5cm}
\includegraphics[width=1\textwidth,height=\textheight]{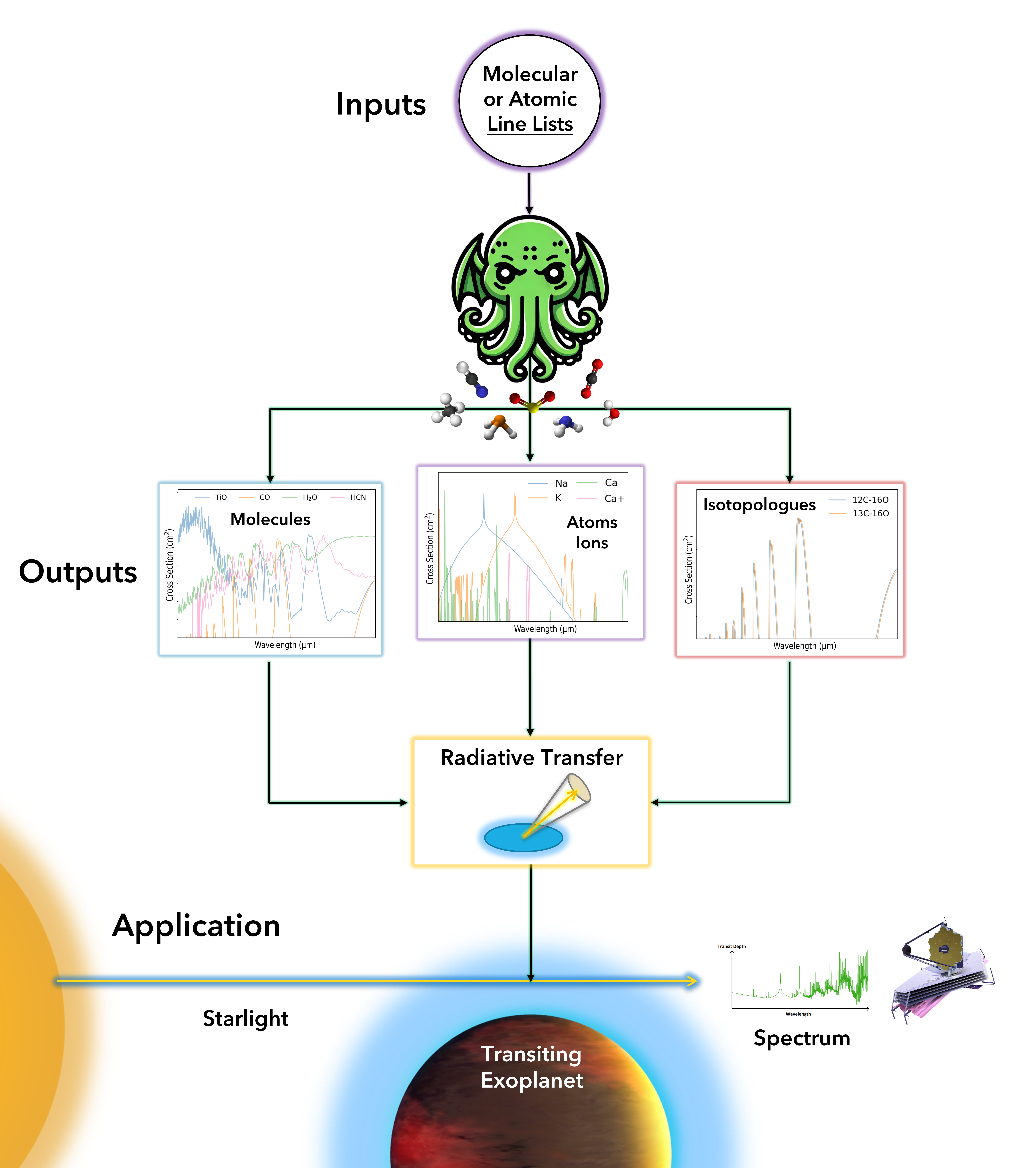}
\caption{The role and applications of the \texttt{Cthulhu} Python package. \texttt{Cthulhu} can download molecular and atomic line lists and calculate the corresponding absorption cross sections as a function of temperature, pressure, and wavenumber. Cross sections made by \texttt{Cthulhu} can be used in radiative transfer codes to calculate model spectra of exoplanet and brown dwarf atmospheres.
\label{fig:Cthulhu_architecture}}
\end{figure}

The purpose of the \texttt{Cthulhu} package is schematically represented in \autoref{fig:Cthulhu_architecture}. Here we walk through this flowchart, highlighting major use cases of \texttt{Cthulhu} and the package's role in the broader process of modelling exoplanetary and brown dwarf atmospheres.

\autoref{fig:Cthulhu_architecture} illustrates three applications of \texttt{Cthulhu}: (i) molecular cross section calculations for hot giant exoplanets; (ii) atomic and ionic cross  sections, including sub-Voigt wings for the Na and K resonance doublets; and (iii) cross sections for different molecular isotopologues.

The first use of \texttt{Cthulhu} is to download existing molecular line lists from online databases. \texttt{Cthulhu}'s \texttt{summon} function can automatically download lines lists from ExoMol and HITRAN/HITEMP and reformat the line lists into space-efficient HDF5 files. Ancillary input files required to calculate cross sections, such as partition functions and pressure broadening files, are also downloaded automatically. Alternatively, the user may manually download a line list from their
respective websites and point \texttt{Cthulhu} to the directory hosting the files. VALD line lists must be downloaded manually by a user with an account on \url{http://vald.astro.uu.se/} (given the terms of use for VALD3), but we provide instructions on how to do this in the \texttt{Cthulhu} documentation. \texttt{Cthulhu} currently supports ExoMol, HITRAN, HITEMP, and VALD line lists, though we welcome user
requests for additional line list database support. Once a line list has been downloaded, the user can move onto the next major use case of \texttt{Cthulhu}, computing cross sections.

The foremost feature of \texttt{Cthulhu} is its ability to straightforwardly compute atomic and molecular cross sections at high speeds (typically \textgreater{} 100,000 lines per second on one CPU).
\texttt{Cthulhu} calculates cross sections via a generalisation of the Vectorised Voigt method (\protect\hyperlink{Yurchenko:2018}{Yurchenko et al., 2018}), whereby our update to the algorithm uses complex derivatives to perturb a grid of pre-computed template Voigt profiles to the specific properties of each given line (see \protect\hyperlink{MacDonald:2019}{MacDonald, 2019}, Chapter 5, for the mathematical description). \texttt{Cthulhu} is accessible, as it does not require GPUs, can run on a standard laptop, and as a pure Python code it is easy for beginners to
install and use. To compute a cross section, a user simply calls \texttt{Cthulhu}'s \texttt{compute\_cross\_section} function, specifying the location of the line list, the temperature and pressure, and the wavenumber range. More advanced users can specify custom settings via optional arguments (e.g.~Voigt wing cutoffs, intensity cutoffs, or a user-provided pressure broadening file). The documentation and function docstrings explain the various arguments users can provide to \texttt{compute\_cross\_section}. The computed cross section is output by default as a .txt file in the \texttt{output} folder on the user's machine, which can be readily used to create an opacity database in a user-preferred format for a specific radiative transfer code. \texttt{Cthulhu} also offers utility functions to combine multiple cross section .txt files (e.g.~a grid of cross sections for different
temperatures and pressures for one or more chemical species) into an HDF5 cross section database, as illustrated in our \href{https://cthulhu.readthedocs.io/en/latest/content/notebooks/quick_start.html}{\texttt{quick\ start}} guide.

The cross section database HDF5 files produced by \texttt{Cthulhu} can be readily adapted for the user's favourite exoplanet or brown dwarf modelling or retrieval code. Cross section HDF5 files calculated by
\texttt{Cthulhu} are already being used in the \texttt{POSEIDON} retrieval code (\protect\hyperlink{MacDonald:2023}{MacDonald, 2023}) and the opacity formats for other retrieval codes will be natively supported soon. The lower part of \autoref{fig:Cthulhu_architecture} illustrates an application of \texttt{Cthulhu}'s cross sections, namely the calculation of exoplanet transmission spectra. Exoplanet transmission spectra modelling codes use cross sections to solve the equation of radiative transfer (e.g. \protect\hyperlink{MacDonald:2022}{MacDonald \& Lewis, 2022}) by calculating the slant optical depth (the integral of the extinction coefficient along the line of sight; see e.g. \protect\hyperlink{Fortney:2005}{Fortney, 2005}) for many rays of light passing through the exoplanet atmosphere. The final output of such a transmission spectrum calculation is the effective planet-star area ratio (i.e.~transit depth) as a function of wavelength seen by a distant observer. Cross sections are similarly used in other observing geometries (e.g.~secondary eclipses of exoplanets or directly imaged brown dwarfs) to calculate the attenuation of light along the path of each light ray.

\hypertarget{future-developments}{%
\section{Future Developments}\label{future-developments}}

\texttt{Cthulhu} v1.0 supports line lists from the commonly used ExoMol, HITRAN, HITEMP, and VALD databases, but support for other databases (e.g.~Kurucz) can be added in the future. \texttt{Cthulhu} currently uses Voigt profiles by default (with the exception of the strong Na and K resonance features), but more complex line profiles (e.g.~speed-dependent Voigt) are under consideration for future releases. We will shortly add additional HDF5 packaging functions to generate out-of-the-box the opacity formats used by other open source exoplanet and brown dwarf retrieval codes, including petitRADTRANS (\protect\hyperlink{Molliere:2019}{Mollière et al., 2019}), TauREx (\protect\hyperlink{Waldmann:2015}{Waldmann et al., 2015}), CHIMERA (\protect\hyperlink{Line:2013}{Line et al., 2013}), and Brewster (\protect\hyperlink{Burningham:2017}{Burningham et al., 2017}). Suggestions for additional features are more than welcome and the easiest way is by \href{https://github.com/MartianColonist/Cthulhu/issues}{\texttt{opening\ an\ issue}} on GitHub. To contribute directly to \texttt{Cthulhu}, please see our \href{https://cthulhu.readthedocs.io/en/latest/content/contributing.html}{\texttt{contribution\ guidelines}}.

\hypertarget{documentation}{%
\section{Documentation}\label{documentation}}

Documentation for \texttt{Cthulhu}, with a quick start guide and step-by-step tutorials, is available at
\url{https://cthulhu.readthedocs.io/en/latest/}.

\hypertarget{similar-tools}{%
\section{Similar Tools}\label{similar-tools}}

\href{https://github.com/exoclime/HELIOS-K}{\texttt{HELIOS-K}} (\protect\hyperlink{Grimm:2015}{Grimm \& Heng, 2015}; \protect\hyperlink{Grimm:2021}{Grimm et al., 2021}),
\href{https://github.com/Trovemaster/exocross}{\texttt{ExoCross}} (\protect\hyperlink{Yurchenko:2018}{Yurchenko et al., 2018}),
\href{https://github.com/radis/radis}{\texttt{RADIS}} (\protect\hyperlink{Pannier:2019}{Pannier \& Laux, 2019}),
\href{https://github.com/Beryl-Jingxin/PyExoCross}{\texttt{PyExoCross}} (\protect\hyperlink{Zhang:2024}{Zhang et al., 2024})

\hypertarget{acknowledgements}{%
\section{Acknowledgements}\label{acknowledgements}}

RJM acknowledges support from NASA through the NASA Hubble Fellowship grant HST-HF2-51513.001 awarded by the Space Telescope Science Institute, which is operated by the Association of Universities for Research in Astronomy, Inc., for NASA, under contract NAS5-26555. RJM thanks Sergei Yurchenko, Mark Marley, Natasha Batalha, Ehsan Gharib-Nezhad, Robert Hargreaves, and Iouli Gordon for helpful discussions on line lists, opacities, and cross section computation techniques.

\hypertarget{references}{%
\section*{References}\label{references}}
\addcontentsline{toc}{section}{References}

\hypertarget{refs}{}
\begin{CSLReferences}{1}{0}
\leavevmode\hypertarget{Anisman:2022}{}%
Anisman, L. O., Chubb, K. L., Changeat, Q., Edwards, B., Yurchenko, S.
N., Tennyson, J., \& Tinetti, G. (2022). {Cross-sections for heavy
atmospheres: H\(_{2}\)O self-broadening}. \emph{Journal of Quantitative
Spectroscopy and Radiative Transfer}, \emph{283}, 108146.
\url{https://doi.org/10.1016/j.jqsrt.2022.108146}

\leavevmode\hypertarget{Barstow:2020}{}%
Barstow, J. K., Changeat, Q., Garland, R., Line, M. R., Rocchetto, M.,
\& Waldmann, I. P. (2020). {A comparison of exoplanet spectroscopic
retrieval tools}. \emph{Monthly Notices of the Royal Astronomical
Society}, \emph{493}(4), 4884--4909.
\url{https://doi.org/10.1093/mnras/staa548}

\leavevmode\hypertarget{Beiler:2023}{}%
Beiler, S. A., Cushing, M. C., Kirkpatrick, J. D., Schneider, A. C.,
Mukherjee, S., \& Marley, M. S. (2023). {The First JWST Spectral Energy
Distribution of a Y Dwarf}. \emph{Astrophysical Journal Letters},
\emph{951}(2), L48. \url{https://doi.org/10.3847/2041-8213/ace32c}

\leavevmode\hypertarget{Birkby:2017}{}%
Birkby, J. L., de Kok, R. J., Brogi, M., Schwarz, H., \& Snellen, I. A.
G. (2017). {Discovery of Water at High Spectral Resolution in the
Atmosphere of 51 Peg b}. \emph{Astronomical Journal}, \emph{153}(3),
138. \url{https://doi.org/10.3847/1538-3881/aa5c87}

\leavevmode\hypertarget{Brogi:2019}{}%
Brogi, M., \& Line, M. R. (2019). {Retrieving Temperatures and
Abundances of Exoplanet Atmospheres with High-resolution
Cross-correlation Spectroscopy}. \emph{Astronomical Journal},
\emph{157}(3), 114. \url{https://doi.org/10.3847/1538-3881/aaffd3}

\leavevmode\hypertarget{Burningham:2017}{}%
Burningham, B., Marley, M. S., Line, M. R., Lupu, R., Visscher, C.,
Morley, C. V., Saumon, D., \& Freedman, R. (2017). {Retrieval of
atmospheric properties of cloudy L dwarfs}. \emph{Monthly Notices of the
Royal Astronomical Society}, \emph{470}(1), 1177--1197.
\url{https://doi.org/10.1093/mnras/stx1246}

\leavevmode\hypertarget{Carter:2024}{}%
Carter, A. L., May, E. M., Espinoza, N., Welbanks, L., Ahrer, E.,
Alderson, L., Brahm, R., Feinstein, A. D., Grant, D., Line, M., Morello,
G., O'Steen, R., Radica, M., Rustamkulov, Z., Stevenson, K. B., Turner,
J. D., Alam, M. K., Anderson, D. R., Batalha, N. M., \ldots{} Zhang, X.
(2024). {A benchmark JWST near-infrared spectrum for the exoplanet
WASP-39 b}. \emph{Nature Astronomy}.
\url{https://doi.org/10.1038/s41550-024-02292-x}

\leavevmode\hypertarget{Fanson:2022}{}%
Fanson, J., Bernstein, R., Ashby, D., Bigelow, B., Brossus, G., Burgett,
W., Demers, R., Fischer, B., Figueroa, F., Groark, F., Laskin, R.,
Millan-Gabet, R., Park, S., Pi, M., Turner, R., \& Walls, B. (2022).
{Overview and status of the Giant Magellan Telescope project}. In H. K.
Marshall, J. Spyromilio, \& T. Usuda (Eds.), \emph{Ground-based and
airborne telescopes IX} (Vol. 12182, p. 121821C).
\url{https://doi.org/10.1117/12.2631694}

\leavevmode\hypertarget{Fortney:2005}{}%
Fortney, J. J. (2005). {The effect of condensates on the
characterization of transiting planet atmospheres with transmission
spectroscopy}. \emph{Monthly Notices of the Royal Astronomical Society},
\emph{364}(2), 649--653.
\url{https://doi.org/10.1111/j.1365-2966.2005.09587.x}

\leavevmode\hypertarget{Gardner:2023}{}%
Gardner, J. P., Mather, J. C., Abbott, R., Abell, J. S., Abernathy, M.,
Abney, F. E., Abraham, J. G., Abraham, R., Abul-Huda, Y. M., Acton, S.,
\& al., et. (2023). {The James Webb Space Telescope Mission}.
\emph{Publications of the Astronomical Society of the Pacific},
\emph{135}(1048), 068001. \url{https://doi.org/10.1088/1538-3873/acd1b5}

\leavevmode\hypertarget{Gharib-Nezhad:2019}{}%
Gharib-Nezhad, E., \& Line, M. R. (2019). {The Influence of H\(_{2}\)O
Pressure Broadening in High-metallicity Exoplanet Atmospheres}.
\emph{Astrophysical Journal}, \emph{872}(1), 27.
\url{https://doi.org/10.3847/1538-4357/aafb7b}

\leavevmode\hypertarget{Gordon:2022}{}%
Gordon, I. E., Rothman, L. S., Hargreaves, R. J., Hashemi, R.,
Karlovets, E. V., Skinner, F. M., Conway, E. K., Hill, C., Kochanov, R.
V., Tan, Y., Wcisło, P., Finenko, A. A., Nelson, K., Bernath, P. F.,
Birk, M., Boudon, V., Campargue, A., Chance, K. V., Coustenis, A.,
\ldots{} Yurchenko, S. N. (2022). {The HITRAN2020 molecular
spectroscopic database}. \emph{Journal of Quantitative Spectroscopy and
Radiative Transfer}, \emph{277}, 107949.
\url{https://doi.org/10.1016/j.jqsrt.2021.107949}

\leavevmode\hypertarget{Grimm:2015}{}%
Grimm, S. L., \& Heng, K. (2015). {HELIOS-K: An Ultrafast, Open-source
Opacity Calculator for Radiative Transfer}. \emph{Astrophysical
Journal}, \emph{808}(2), 182.
\url{https://doi.org/10.1088/0004-637X/808/2/182}

\leavevmode\hypertarget{Grimm:2021}{}%
Grimm, S. L., Malik, M., Kitzmann, D., Guzmán-Mesa, A., Hoeijmakers, H.
J., Fisher, C., Mendonça, J. M., Yurchenko, S. N., Tennyson, J.,
Alesina, F., Buchschacher, N., Burnier, J., Segransan, D., Kurucz, R.
L., \& Heng, K. (2021). {HELIOS-K 2.0 Opacity Calculator and Open-source
Opacity Database for Exoplanetary Atmospheres}. \emph{Astrophysical
Journal Supplement}, \emph{253}(1), 30.
\url{https://doi.org/10.3847/1538-4365/abd773}

\leavevmode\hypertarget{Hedges:2016}{}%
Hedges, C., \& Madhusudhan, N. (2016). {Effect of pressure broadening on
molecular absorption cross sections in exoplanetary atmospheres}.
\emph{Monthly Notices of the Royal Astronomical Society}, \emph{458}(2),
1427--1449. \url{https://doi.org/10.1093/mnras/stw278}

\leavevmode\hypertarget{Heng:2017}{}%
Heng, K. (2017). \emph{{Exoplanetary Atmospheres: Theoretical Concepts
and Foundations}}.

\leavevmode\hypertarget{Line:2013}{}%
Line, M. R., Wolf, A. S., Zhang, X., Knutson, H., Kammer, J. A.,
Ellison, E., Deroo, P., Crisp, D., \& Yung, Y. L. (2013). {A Systematic
Retrieval Analysis of Secondary Eclipse Spectra. I. A Comparison of
Atmospheric Retrieval Techniques}. \emph{Astrophysical Journal},
\emph{775}(2), 137. \url{https://doi.org/10.1088/0004-637X/775/2/137}

\leavevmode\hypertarget{MacDonald:2019}{}%
MacDonald. (2019). \emph{{Revealing the nature of exoplanetary
atmospheres}} {[}PhD thesis{]}. University of Cambridge, UK. \url{https://doi.org/10.17863/CAM.44898}

\leavevmode\hypertarget{MacDonald:2023}{}%
MacDonald. (2023). {POSEIDON: A Multidimensional Atmospheric Retrieval
Code for Exoplanet Spectra}. \emph{The Journal of Open Source Software},
\emph{8}, 4873. \url{https://doi.org/10.21105/joss.04873}

\leavevmode\hypertarget{MacDonald:2022}{}%
MacDonald, \& Lewis. (2022). {TRIDENT: A Rapid 3D Radiative-transfer
Model for Exoplanet Transmission Spectra}. \emph{Astrophysical Journal},
\emph{929}(1), 20. \url{https://doi.org/10.3847/1538-4357/ac47fe}

\leavevmode\hypertarget{Merritt:2020}{}%
Merritt, S. R., Gibson, N. P., Nugroho, S. K., de Mooij, E. J. W.,
Hooton, M. J., Matthews, S. M., McKemmish, L. K., Mikal-Evans, T.,
Nikolov, N., Sing, D. K., Spake, J. J., \& Watson, C. A. (2020).
{Non-detection of TiO and VO in the atmosphere of WASP-121b using
high-resolution spectroscopy}. \emph{Astronomy \& Astrophysics},
\emph{636}, A117. \url{https://doi.org/10.1051/0004-6361/201937409}

\leavevmode\hypertarget{Miles:2023}{}%
Miles, B. E., Biller, B. A., Patapis, P., Worthen, K., Rickman, E.,
Hoch, K. K. W., Skemer, A., Perrin, M. D., Whiteford, N., Chen, C. H.,
Sargent, B., Mukherjee, S., Morley, C. V., Moran, S. E., Bonnefoy, M.,
Petrus, S., Carter, A. L., Choquet, E., Hinkley, S., \ldots{} Zhang, Z.
(2023). {The JWST Early-release Science Program for Direct Observations
of Exoplanetary Systems II: A 1 to 20 {\(\mu\)}m Spectrum of the
Planetary-mass Companion VHS 1256-1257 b}. \emph{Astrophysical Journal
Letters}, \emph{946}(1), L6.
\url{https://doi.org/10.3847/2041-8213/acb04a}

\leavevmode\hypertarget{Molliere:2019}{}%
Mollière, P., Wardenier, J. P., van Boekel, R., Henning, Th.,
Molaverdikhani, K., \& Snellen, I. A. G. (2019). {petitRADTRANS. A
Python radiative transfer package for exoplanet characterization and
retrieval}. \emph{Astronomy \& Astrophysics}, \emph{627}, A67.
\url{https://doi.org/10.1051/0004-6361/201935470}

\leavevmode\hypertarget{Padovani:2023}{}%
Padovani, P., \& Cirasuolo, M. (2023). {The Extremely Large Telescope}.
\emph{Contemporary Physics}, \emph{64}(1), 47--64.
\url{https://doi.org/10.1080/00107514.2023.2266921}

\leavevmode\hypertarget{Pakhomov:2017}{}%
Pakhomov, Yu., Piskunov, N., \& Ryabchikova, T. (2017). {VALD3: Current
Developments}. In Yu. Yu. Balega, D. O. Kudryavtsev, I. I. Romanyuk, \&
I. A. Yakunin (Eds.), \emph{Stars: From collapse to collapse} (Vol. 510,
p. 518). \url{https://doi.org/10.48550/arXiv.1710.10854}

\leavevmode\hypertarget{Pannier:2019}{}%
Pannier, E., \& Laux, C. O. (2019). {RADIS: A nonequilibrium
line-by-line radiative code for CO\(_{2}\) and HITRAN-like database
species}. \emph{Journal of Quantitative Spectroscopy and Radiative
Transfer}, \emph{222}, 12--25.
\url{https://doi.org/10.1016/j.jqsrt.2018.09.027}

\leavevmode\hypertarget{Pelletier:2023}{}%
Pelletier, S., Benneke, B., Ali-Dib, M., Prinoth, B., Kasper, D.,
Seifahrt, A., Bean, J. L., Debras, F., Klein, B., Bazinet, L.,
Hoeijmakers, H. J., Kesseli, A. Y., Lim, O., Carmona, A., Pino, L.,
Casasayas-Barris, N., Hood, T., \& Stürmer, J. (2023). {Vanadium oxide
and a sharp onset of cold-trapping on a giant exoplanet}. \emph{Nature},
\emph{619}(7970), 491--494.
\url{https://doi.org/10.1038/s41586-023-06134-0}

\leavevmode\hypertarget{Rothman:2010}{}%
Rothman, L. S., Gordon, I. E., Barber, R. J., Dothe, H., Gamache, R. R.,
Goldman, A., Perevalov, V. I., Tashkun, S. A., \& Tennyson, J. (2010).
{HITEMP, the high-temperature molecular spectroscopic database}.
\emph{Journal of Quantitative Spectroscopy and Radiative Transfer},
\emph{111}, 2139--2150.
\url{https://doi.org/10.1016/j.jqsrt.2010.05.001}

\leavevmode\hypertarget{Seager:2010}{}%
Seager, S. (2010). \emph{{Exoplanet Atmospheres: Physical Processes}}.

\leavevmode\hypertarget{Skidmore:2015}{}%
Skidmore, W., TMT International Science Development Teams, \& Science
Advisory Committee, T. (2015). {Thirty Meter Telescope Detailed Science
Case: 2015}. \emph{Research in Astronomy and Astrophysics},
\emph{15}(12), 1945. \url{https://doi.org/10.1088/1674-4527/15/12/001}

\leavevmode\hypertarget{Snellen:2010}{}%
Snellen, I. A. G., de Kok, R. J., de Mooij, E. J. W., \& Albrecht, S.
(2010). {The orbital motion, absolute mass and high-altitude winds of
exoplanet HD209458b}. \emph{Nature}, \emph{465}(7301), 1049--1051.
\url{https://doi.org/10.1038/nature09111}

\leavevmode\hypertarget{Tennyson:2020}{}%
Tennyson, J., Yurchenko, S. N., Al-Refaie, A. F., Clark, V. H. J.,
Chubb, K. L., Conway, E. K., Dewan, A., Gorman, M. N., Hill, C.,
Lynas-Gray, A. E., Mellor, T., McKemmish, L. K., Owens, A., Polyansky,
O. L., Semenov, M., Somogyi, W., Tinetti, G., Upadhyay, A., Waldmann,
I., \ldots{} Yurchenko, O. P. (2020). {The 2020 release of the ExoMol
database: Molecular line lists for exoplanet and other hot atmospheres}.
\emph{Journal of Quantitative Spectroscopy and Radiative Transfer},
\emph{255}, 107228. \url{https://doi.org/10.1016/j.jqsrt.2020.107228}

\leavevmode\hypertarget{Waldmann:2015}{}%
Waldmann, I. P., Tinetti, G., Rocchetto, M., Barton, E. J., Yurchenko,
S. N., \& Tennyson, J. (2015). {Tau-REx I: A Next Generation Retrieval
Code for Exoplanetary Atmospheres}. \emph{Astrophysical Journal},
\emph{802}(2), 107. \url{https://doi.org/10.1088/0004-637X/802/2/107}

\leavevmode\hypertarget{Welbanks:2024}{}%
Welbanks, L., Bell, T. J., Beatty, T. G., Line, M. R., Ohno, K.,
Fortney, J. J., Schlawin, E., Greene, T. P., Rauscher, E., McGill, P.,
Murphy, M., Parmentier, V., Tang, Y., Edelman, I., Mukherjee, S., Wiser,
L. S., Lagage, P.-O., Dyrek, A., \& Arnold, K. E. (2024). {A high
internal heat flux and large core in a warm Neptune exoplanet}.
\emph{Nature}, \emph{630}(8018), 836--840.
\url{https://doi.org/10.1038/s41586-024-07514-w}

\leavevmode\hypertarget{Yurchenko:2018}{}%
Yurchenko, S. N., Al-Refaie, A. F., \& Tennyson, J. (2018). {EXOCROSS: a
general program for generating spectra from molecular line lists}.
\emph{Astronomy \& Astrophysics}, \emph{614}, A131.
\url{https://doi.org/10.1051/0004-6361/201732531}

\leavevmode\hypertarget{Zhang:2024}{}%
Zhang, J., Tennyson, J., \& Yurchenko, S. N. (2024). {PYEXOCROSS: a
Python program for generating spectra and cross-sections from molecular
line lists}. \emph{RAS Techniques and Instruments}, \emph{3}(1),
257--287. \url{https://doi.org/10.1093/rasti/rzae016}

\end{CSLReferences}


@ARTICLE{Tennyson:2020,
       author = {{Tennyson}, Jonathan and {Yurchenko}, Sergei N. and {Al-Refaie}, Ahmed F. and {Clark}, Victoria H.~J. and {Chubb}, Katy L. and {Conway}, Eamon K. and {Dewan}, Akhil and {Gorman}, Maire N. and {Hill}, Christian and {Lynas-Gray}, A.~E. and {Mellor}, Thomas and {McKemmish}, Laura K. and {Owens}, Alec and {Polyansky}, Oleg L. and {Semenov}, Mikhail and {Somogyi}, Wilfrid and {Tinetti}, Giovanna and {Upadhyay}, Apoorva and {Waldmann}, Ingo and {Wang}, Yixin and {Wright}, Samuel and {Yurchenko}, Olga P.},
        title = "{The 2020 release of the ExoMol database: Molecular line lists for exoplanet and other hot atmospheres}",
      journal = {Journal of Quantitative Spectroscopy and Radiative Transfer},
         year = 2020,
        month = nov,
       volume = {255},
          eid = {107228},
        pages = {107228},
          doi = {10.1016/j.jqsrt.2020.107228},
archivePrefix = {arXiv},
       eprint = {2007.13022},
 primaryClass = {astro-ph.SR},
       adsurl = {https://ui.adsabs.harvard.edu/abs/2020JQSRT.25507228T}
}

@ARTICLE{Rothman:2010,
       author = {{Rothman}, L.~S. and {Gordon}, I.~E. and {Barber}, R.~J. and {Dothe}, H. and {Gamache}, R.~R. and {Goldman}, A. and {Perevalov}, V.~I. and {Tashkun}, S.~A. and {Tennyson}, J.},
        title = "{HITEMP, the high-temperature molecular spectroscopic database}",
      journal = {Journal of Quantitative Spectroscopy and Radiative Transfer},
         year = 2010,
        month = oct,
       volume = {111},
        pages = {2139-2150},
          doi = {10.1016/j.jqsrt.2010.05.001},
       adsurl = {https://ui.adsabs.harvard.edu/abs/2010JQSRT.111.2139R}
}

@ARTICLE{Gordon:2022,
       author = {{Gordon}, I.~E. and {Rothman}, L.~S. and {Hargreaves}, R.~J. and {Hashemi}, R. and {Karlovets}, E.~V. and {Skinner}, F.~M. and {Conway}, E.~K. and {Hill}, C. and {Kochanov}, R.~V. and {Tan}, Y. and {Wcis{\l}o}, P. and {Finenko}, A.~A. and {Nelson}, K. and {Bernath}, P.~F. and {Birk}, M. and {Boudon}, V. and {Campargue}, A. and {Chance}, K.~V. and {Coustenis}, A. and {Drouin}, B.~J. and {Flaud}, J. -M. and {Gamache}, R.~R. and {Hodges}, J.~T. and {Jacquemart}, D. and {Mlawer}, E.~J. and {Nikitin}, A.~V. and {Perevalov}, V.~I. and {Rotger}, M. and {Tennyson}, J. and {Toon}, G.~C. and {Tran}, H. and {Tyuterev}, V.~G. and {Adkins}, E.~M. and {Baker}, A. and {Barbe}, A. and {Can{\`e}}, E. and {Cs{\'a}sz{\'a}r}, A.~G. and {Dudaryonok}, A. and {Egorov}, O. and {Fleisher}, A.~J. and {Fleurbaey}, H. and {Foltynowicz}, A. and {Furtenbacher}, T. and {Harrison}, J.~J. and {Hartmann}, J. -M. and {Horneman}, V. -M. and {Huang}, X. and {Karman}, T. and {Karns}, J. and {Kassi}, S. and {Kleiner}, I. and {Kofman}, V. and {Kwabia-Tchana}, F. and {Lavrentieva}, N.~N. and {Lee}, T.~J. and {Long}, D.~A. and {Lukashevskaya}, A.~A. and {Lyulin}, O.~M. and {Makhnev}, V. Yu. and {Matt}, W. and {Massie}, S.~T. and {Melosso}, M. and {Mikhailenko}, S.~N. and {Mondelain}, D. and {M{\"u}ller}, H.~S.~P. and {Naumenko}, O.~V. and {Perrin}, A. and {Polyansky}, O.~L. and {Raddaoui}, E. and {Raston}, P.~L. and {Reed}, Z.~D. and {Rey}, M. and {Richard}, C. and {T{\'o}bi{\'a}s}, R. and {Sadiek}, I. and {Schwenke}, D.~W. and {Starikova}, E. and {Sung}, K. and {Tamassia}, F. and {Tashkun}, S.~A. and {Vander Auwera}, J. and {Vasilenko}, I.~A. and {Vigasin}, A.~A. and {Villanueva}, G.~L. and {Vispoel}, B. and {Wagner}, G. and {Yachmenev}, A. and {Yurchenko}, S.~N.},
        title = "{The HITRAN2020 molecular spectroscopic database}",
      journal = {Journal of Quantitative Spectroscopy and Radiative Transfer},
         year = 2022,
        month = jan,
       volume = {277},
          eid = {107949},
        pages = {107949},
          doi = {10.1016/j.jqsrt.2021.107949},
       adsurl = {https://ui.adsabs.harvard.edu/abs/2022JQSRT.27707949G}
}

@INPROCEEDINGS{Pakhomov:2017,
       author = {{Pakhomov}, Yu. and {Piskunov}, N. and {Ryabchikova}, T.},
        title = "{VALD3: Current Developments}",
    booktitle = {Stars: From Collapse to Collapse},
         year = 2017,
       editor = {{Balega}, Yu. Yu. and {Kudryavtsev}, D.~O. and {Romanyuk}, I.~I. and {Yakunin}, I.~A.},
       series = {Astronomical Society of the Pacific Conference Series},
       volume = {510},
        month = jun,
        pages = {518},
          doi = {10.48550/arXiv.1710.10854},
archivePrefix = {arXiv},
       eprint = {1710.10854},
 primaryClass = {astro-ph.IM},
       adsurl = {https://ui.adsabs.harvard.edu/abs/2017ASPC..510..518P}
}

@ARTICLE{Grimm:2015,
       author = {{Grimm}, Simon L. and {Heng}, Kevin},
        title = "{HELIOS-K: An Ultrafast, Open-source Opacity Calculator for Radiative Transfer}",
      journal = {Astrophysical Journal},
         year = 2015,
        month = aug,
       volume = {808},
       number = {2},
          eid = {182},
        pages = {182},
          doi = {10.1088/0004-637X/808/2/182},
archivePrefix = {arXiv},
       eprint = {1503.03806},
 primaryClass = {astro-ph.EP},
       adsurl = {https://ui.adsabs.harvard.edu/abs/2015ApJ...808..182G}
}

@ARTICLE{Grimm:2021,
       author = {{Grimm}, Simon L. and {Malik}, Matej and {Kitzmann}, Daniel and {Guzm{\'a}n-Mesa}, Andrea and {Hoeijmakers}, H. Jens and {Fisher}, Chloe and {Mendon{\c{c}}a}, Jo{\~a}o M. and {Yurchenko}, Sergey N. and {Tennyson}, Jonathan and {Alesina}, Fabien and {Buchschacher}, Nicolas and {Burnier}, Julien and {Segransan}, Damien and {Kurucz}, Robert L. and {Heng}, Kevin},
        title = "{HELIOS-K 2.0 Opacity Calculator and Open-source Opacity Database for Exoplanetary Atmospheres}",
      journal = {Astrophysical Journal Supplement},
         year = 2021,
        month = mar,
       volume = {253},
       number = {1},
          eid = {30},
        pages = {30},
          doi = {10.3847/1538-4365/abd773},
archivePrefix = {arXiv},
       eprint = {2101.02005},
 primaryClass = {astro-ph.EP},
       adsurl = {https://ui.adsabs.harvard.edu/abs/2021ApJS..253...30G}
}

@ARTICLE{Yurchenko:2018,
       author = {{Yurchenko}, Sergei N. and {Al-Refaie}, Ahmed F. and {Tennyson}, Jonathan},
        title = "{EXOCROSS: a general program for generating spectra from molecular line lists}",
      journal = {Astronomy \& Astrophysics},
         year = 2018,
        month = jun,
       volume = {614},
          eid = {A131},
        pages = {A131},
          doi = {10.1051/0004-6361/201732531},
archivePrefix = {arXiv},
       eprint = {1801.09803},
 primaryClass = {astro-ph.EP},
       adsurl = {https://ui.adsabs.harvard.edu/abs/2018A&A...614A.131Y}
}

@ARTICLE{Pannier:2019,
       author = {{Pannier}, Erwan and {Laux}, Christophe O.},
        title = "{RADIS: A nonequilibrium line-by-line radiative code for CO$_{2}$ and HITRAN-like database species}",
      journal = {Journal of Quantitative Spectroscopy and Radiative Transfer},
         year = 2019,
        month = jan,
       volume = {222},
        pages = {12-25},
          doi = {10.1016/j.jqsrt.2018.09.027},
       adsurl = {https://ui.adsabs.harvard.edu/abs/2019JQSRT.222...12P}
}

@ARTICLE{Hedges:2016,
       author = {{Hedges}, Christina and {Madhusudhan}, Nikku},
        title = "{Effect of pressure broadening on molecular absorption cross sections in exoplanetary atmospheres}",
      journal = {Monthly Notices of the Royal Astronomical Society},
         year = 2016,
        month = may,
       volume = {458},
       number = {2},
        pages = {1427-1449},
          doi = {10.1093/mnras/stw278},
archivePrefix = {arXiv},
       eprint = {1602.00751},
 primaryClass = {astro-ph.EP},
       adsurl = {https://ui.adsabs.harvard.edu/abs/2016MNRAS.458.1427H}
}

@ARTICLE{Gharib-Nezhad:2019,
       author = {{Gharib-Nezhad}, Ehsan and {Line}, Michael R.},
        title = "{The Influence of H$_{2}$O Pressure Broadening in High-metallicity Exoplanet Atmospheres}",
      journal = {Astrophysical Journal},
         year = 2019,
        month = feb,
       volume = {872},
       number = {1},
          eid = {27},
        pages = {27},
          doi = {10.3847/1538-4357/aafb7b},
archivePrefix = {arXiv},
       eprint = {1809.02548},
 primaryClass = {astro-ph.EP},
       adsurl = {https://ui.adsabs.harvard.edu/abs/2019ApJ...872...27G}
}

@ARTICLE{Anisman:2022,
       author = {{Anisman}, Lara O. and {Chubb}, Katy L. and {Changeat}, Quentin and {Edwards}, Billy and {Yurchenko}, Sergei N. and {Tennyson}, Jonathan and {Tinetti}, Giovanna},
        title = "{Cross-sections for heavy atmospheres: H$_{2}$O self-broadening}",
      journal = {Journal of Quantitative Spectroscopy and Radiative Transfer},
         year = 2022,
        month = jun,
       volume = {283},
          eid = {108146},
        pages = {108146},
          doi = {10.1016/j.jqsrt.2022.108146},
archivePrefix = {arXiv},
       eprint = {2203.02335},
 primaryClass = {astro-ph.EP},
       adsurl = {https://ui.adsabs.harvard.edu/abs/2022JQSRT.28308146A}
}

@ARTICLE{MacDonald:2022,
       author = {{MacDonald} and {Lewis}},
        title = "{TRIDENT: A Rapid 3D Radiative-transfer Model for Exoplanet Transmission Spectra}",
      journal = {Astrophysical Journal},
         year = 2022,
        month = apr,
       volume = {929},
       number = {1},
          eid = {20},
        pages = {20},
          doi = {10.3847/1538-4357/ac47fe},
archivePrefix = {arXiv},
       eprint = {2111.05862},
 primaryClass = {astro-ph.EP},
       adsurl = {https://ui.adsabs.harvard.edu/abs/2022ApJ...929...20M}
}

@ARTICLE{MacDonald:2023,
       author = {{MacDonald}},
        title = "{POSEIDON: A Multidimensional Atmospheric Retrieval Code for Exoplanet Spectra}",
      journal = {The Journal of Open Source Software},
         year = 2023,
        month = jan,
       volume = {8},
          eid = {4873},
        pages = {4873},
          doi = {10.21105/joss.04873},
       adsurl = {https://ui.adsabs.harvard.edu/abs/2023JOSS....8.4873M}
}

@ARTICLE{Molliere:2019,
       author = {{Molli{\`e}re}, P. and {Wardenier}, J.~P. and {van Boekel}, R. and {Henning}, Th. and {Molaverdikhani}, K. and {Snellen}, I.~A.~G.},
        title = "{petitRADTRANS. A Python radiative transfer package for exoplanet characterization and retrieval}",
      journal = {Astronomy \& Astrophysics},
         year = 2019,
        month = jul,
       volume = {627},
          eid = {A67},
        pages = {A67},
          doi = {10.1051/0004-6361/201935470},
archivePrefix = {arXiv},
       eprint = {1904.11504},
 primaryClass = {astro-ph.EP},
       adsurl = {https://ui.adsabs.harvard.edu/abs/2019A&A...627A..67M}
}

@ARTICLE{Waldmann:2015,
       author = {{Waldmann}, I.~P. and {Tinetti}, G. and {Rocchetto}, M. and {Barton}, E.~J. and {Yurchenko}, S.~N. and {Tennyson}, J.},
        title = "{Tau-REx I: A Next Generation Retrieval Code for Exoplanetary Atmospheres}",
      journal = {Astrophysical Journal},
         year = 2015,
        month = apr,
       volume = {802},
       number = {2},
          eid = {107},
        pages = {107},
          doi = {10.1088/0004-637X/802/2/107},
archivePrefix = {arXiv},
       eprint = {1409.2312},
 primaryClass = {astro-ph.EP},
       adsurl = {https://ui.adsabs.harvard.edu/abs/2015ApJ...802..107W}
}

@ARTICLE{Zhang:2024,
       author = {{Zhang}, Jingxin and {Tennyson}, Jonathan and {Yurchenko}, Sergei N},
        title = "{PYEXOCROSS: a Python program for generating spectra and cross-sections from molecular line lists}",
      journal = {RAS Techniques and Instruments},
         year = 2024,
        month = jan,
       volume = {3},
       number = {1},
        pages = {257-287},
          doi = {10.1093/rasti/rzae016},
archivePrefix = {arXiv},
       eprint = {2406.03977},
 primaryClass = {astro-ph.IM},
       adsurl = {https://ui.adsabs.harvard.edu/abs/2024RASTI...3..257Z}
}

@ARTICLE{Merritt:2020,
       author = {{Merritt}, S.~R. and {Gibson}, N.~P. and {Nugroho}, S.~K. and {de Mooij}, E.~J.~W. and {Hooton}, M.~J. and {Matthews}, S.~M. and {McKemmish}, L.~K. and {Mikal-Evans}, T. and {Nikolov}, N. and {Sing}, D.~K. and {Spake}, J.~J. and {Watson}, C.~A.},
        title = "{Non-detection of TiO and VO in the atmosphere of WASP-121b using high-resolution spectroscopy}",
      journal = {Astronomy \& Astrophysics},
         year = 2020,
        month = apr,
       volume = {636},
          eid = {A117},
        pages = {A117},
          doi = {10.1051/0004-6361/201937409},
archivePrefix = {arXiv},
       eprint = {2002.02795},
 primaryClass = {astro-ph.EP},
       adsurl = {https://ui.adsabs.harvard.edu/abs/2020A&A...636A.117M}
}

@ARTICLE{Brogi:2019,
       author = {{Brogi}, Matteo and {Line}, Michael R.},
        title = "{Retrieving Temperatures and Abundances of Exoplanet Atmospheres with High-resolution Cross-correlation Spectroscopy}",
      journal = {Astronomical Journal},
         year = 2019,
        month = mar,
       volume = {157},
       number = {3},
          eid = {114},
        pages = {114},
          doi = {10.3847/1538-3881/aaffd3},
archivePrefix = {arXiv},
       eprint = {1811.01681},
 primaryClass = {astro-ph.EP},
       adsurl = {https://ui.adsabs.harvard.edu/abs/2019AJ....157..114B}
}

@ARTICLE{Miles:2023,
       author = {{Miles}, Brittany E. and {Biller}, Beth A. and {Patapis}, Polychronis and {Worthen}, Kadin and {Rickman}, Emily and {Hoch}, Kielan K.~W. and {Skemer}, Andrew and {Perrin}, Marshall D. and {Whiteford}, Niall and {Chen}, Christine H. and {Sargent}, B. and {Mukherjee}, Sagnick and {Morley}, Caroline V. and {Moran}, Sarah E. and {Bonnefoy}, Mickael and {Petrus}, Simon and {Carter}, Aarynn L. and {Choquet}, Elodie and {Hinkley}, Sasha and {Ward-Duong}, Kimberly and {Leisenring}, Jarron M. and {Millar-Blanchaer}, Maxwell A. and {Pueyo}, Laurent and {Ray}, Shrishmoy and {Sallum}, Steph and {Stapelfeldt}, Karl R. and {Stone}, Jordan M. and {Wang}, Jason J. and {Absil}, Olivier and {Balmer}, William O. and {Boccaletti}, Anthony and {Bonavita}, Mariangela and {Booth}, Mark and {Bowler}, Brendan P. and {Chauvin}, Gael and {Christiaens}, Valentin and {Currie}, Thayne and {Danielski}, Camilla and {Fortney}, Jonathan J. and {Girard}, Julien H. and {Grady}, Carol A. and {Greenbaum}, Alexandra Z. and {Henning}, Thomas and {Hines}, Dean C. and {Janson}, Markus and {Kalas}, Paul and {Kammerer}, Jens and {Kennedy}, Grant M. and {Kenworthy}, Matthew A. and {Kervella}, Pierre and {Lagage}, Pierre-Olivier and {Lew}, Ben W.~P. and {Liu}, Michael C. and {Macintosh}, Bruce and {Marino}, Sebastian and {Marley}, Mark S. and {Marois}, Christian and {Matthews}, Elisabeth C. and {Matthews}, Brenda C. and {Mawet}, Dimitri and {McElwain}, Michael W. and {Metchev}, Stanimir and {Meyer}, Michael R. and {Molliere}, Paul and {Pantin}, Eric and {Quirrenbach}, Andreas and {Rebollido}, Isabel and {Ren}, Bin B. and {Schneider}, Glenn and {Vasist}, Malavika and {Wyatt}, Mark C. and {Zhou}, Yifan and {Briesemeister}, Zackery W. and {Bryan}, Marta L. and {Calissendorff}, Per and {Cantalloube}, Faustine and {Cugno}, Gabriele and {De Furio}, Matthew and {Dupuy}, Trent J. and {Factor}, Samuel M. and {Faherty}, Jacqueline K. and {Fitzgerald}, Michael P. and {Franson}, Kyle and {Gonzales}, Eileen C. and {Hood}, Callie E. and {Howe}, Alex R. and {Kraus}, Adam L. and {Kuzuhara}, Masayuki and {Lagrange}, Anne-Marie and {Lawson}, Kellen and {Lazzoni}, Cecilia and {Liu}, Pengyu and {Llop-Sayson}, Jorge and {Lloyd}, James P. and {Martinez}, Raquel A. and {Mazoyer}, Johan and {Quanz}, Sascha P. and {Redai}, Jea Adams and {Samland}, Matthias and {Schlieder}, Joshua E. and {Tamura}, Motohide and {Tan}, Xianyu and {Uyama}, Taichi and {Vigan}, Arthur and {Vos}, Johanna M. and {Wagner}, Kevin and {Wolff}, Schuyler G. and {Ygouf}, Marie and {Zhang}, Xi and {Zhang}, Keming and {Zhang}, Zhoujian},
        title = "{The JWST Early-release Science Program for Direct Observations of Exoplanetary Systems II: A 1 to 20 {\ensuremath{\mu}}m Spectrum of the Planetary-mass Companion VHS 1256-1257 b}",
      journal = {Astrophysical Journal Letters},
         year = 2023,
        month = mar,
       volume = {946},
       number = {1},
          eid = {L6},
        pages = {L6},
          doi = {10.3847/2041-8213/acb04a},
archivePrefix = {arXiv},
       eprint = {2209.00620},
 primaryClass = {astro-ph.EP},
       adsurl = {https://ui.adsabs.harvard.edu/abs/2023ApJ...946L...6M}
}

@ARTICLE{Beiler:2023,
       author = {{Beiler}, Samuel A. and {Cushing}, Michael C. and {Kirkpatrick}, J. Davy and {Schneider}, Adam C. and {Mukherjee}, Sagnick and {Marley}, Mark S.},
        title = "{The First JWST Spectral Energy Distribution of a Y Dwarf}",
      journal = {Astrophysical Journal Letters},
         year = 2023,
        month = jul,
       volume = {951},
       number = {2},
          eid = {L48},
        pages = {L48},
          doi = {10.3847/2041-8213/ace32c},
archivePrefix = {arXiv},
       eprint = {2306.11807},
 primaryClass = {astro-ph.SR},
       adsurl = {https://ui.adsabs.harvard.edu/abs/2023ApJ...951L..48B}
}

@ARTICLE{Carter:2024,
       author = {{Carter}, A.~L. and {May}, E.~M. and {Espinoza}, N. and {Welbanks}, L. and {Ahrer}, E. and {Alderson}, L. and {Brahm}, R. and {Feinstein}, A.~D. and {Grant}, D. and {Line}, M. and {Morello}, G. and {O'Steen}, R. and {Radica}, M. and {Rustamkulov}, Z. and {Stevenson}, K.~B. and {Turner}, J.~D. and {Alam}, M.~K. and {Anderson}, D.~R. and {Batalha}, N.~M. and {Battley}, M.~P. and {Bayliss}, D. and {Bean}, J.~L. and {Benneke}, B. and {Berta-Thompson}, Z.~K. and {Brande}, J. and {Bryant}, E.~M. and {Burleigh}, M.~R. and {Coulombe}, L. and {Crossfield}, I.~J.~M. and {Damiano}, M. and {D{\'e}sert}, J. -M. and {Flagg}, L. and {Gill}, S. and {Inglis}, J. and {Kirk}, J. and {Knutson}, H. and {Kreidberg}, L. and {L{\'o}pez Morales}, M. and {Mansfield}, M. and {Moran}, S.~E. and {Murray}, C.~A. and {Nixon}, M.~C. and {Petit dit de la Roche}, D.~J.~M. and {Rackham}, B.~V. and {Schlawin}, E. and {Sing}, D.~K. and {Wakeford}, H.~R. and {Wallack}, N.~L. and {Wheatley}, P.~J. and {Zieba}, S. and {Aggarwal}, K. and {Barstow}, J.~K. and {Bell}, T.~J. and {Blecic}, J. and {Caceres}, C. and {Crouzet}, N. and {Cubillos}, P.~E. and {Daylan}, T. and {de Val-Borro}, M. and {Decin}, L. and {Fortney}, J.~J. and {Gibson}, N.~P. and {Heng}, K. and {Hu}, R. and {Kempton}, E.~M. -R. and {Lagage}, P. and {Lothringer}, J.~D. and {Lustig-Yaeger}, J. and {Mancini}, L. and {Mayne}, N.~J. and {Mayorga}, L.~C. and {Molaverdikhani}, K. and {Nasedkin}, E. and {Ohno}, K. and {Parmentier}, V. and {Powell}, D. and {Redfield}, S. and {Roy}, P. and {Taylor}, J. and {Zhang}, X.},
        title = "{A benchmark JWST near-infrared spectrum for the exoplanet WASP-39 b}",
      journal = {Nature Astronomy},
         year = 2024,
        month = jul,
          doi = {10.1038/s41550-024-02292-x},
archivePrefix = {arXiv},
       eprint = {2407.13893},
 primaryClass = {astro-ph.EP},
       adsurl = {https://ui.adsabs.harvard.edu/abs/2024NatAs.tmp..128C}
}

@ARTICLE{Welbanks:2024,
       author = {{Welbanks}, Luis and {Bell}, Taylor J. and {Beatty}, Thomas G. and {Line}, Michael R. and {Ohno}, Kazumasa and {Fortney}, Jonathan J. and {Schlawin}, Everett and {Greene}, Thomas P. and {Rauscher}, Emily and {McGill}, Peter and {Murphy}, Matthew and {Parmentier}, Vivien and {Tang}, Yao and {Edelman}, Isaac and {Mukherjee}, Sagnick and {Wiser}, Lindsey S. and {Lagage}, Pierre-Olivier and {Dyrek}, Achr{\`e}ne and {Arnold}, Kenneth E.},
        title = "{A high internal heat flux and large core in a warm Neptune exoplanet}",
      journal = {Nature},
         year = 2024,
        month = jun,
       volume = {630},
       number = {8018},
        pages = {836-840},
          doi = {10.1038/s41586-024-07514-w},
archivePrefix = {arXiv},
       eprint = {2405.11018},
 primaryClass = {astro-ph.EP},
       adsurl = {https://ui.adsabs.harvard.edu/abs/2024Natur.630..836W}
}

@ARTICLE{Barstow:2020,
       author = {{Barstow}, Joanna K. and {Changeat}, Quentin and {Garland}, Ryan and {Line}, Michael R. and {Rocchetto}, Marco and {Waldmann}, Ingo P.},
        title = "{A comparison of exoplanet spectroscopic retrieval tools}",
      journal = {Monthly Notices of the Royal Astronomical Society},
         year = 2020,
        month = apr,
       volume = {493},
       number = {4},
        pages = {4884-4909},
          doi = {10.1093/mnras/staa548},
archivePrefix = {arXiv},
       eprint = {2002.01063},
 primaryClass = {astro-ph.EP},
       adsurl = {https://ui.adsabs.harvard.edu/abs/2020MNRAS.493.4884B}
}

@BOOK{Seager:2010,
       author = {{Seager}, Sara},
        title = "{Exoplanet Atmospheres: Physical Processes}",
         year = 2010,
       adsurl = {https://ui.adsabs.harvard.edu/abs/2010eapp.book.....S}
}

@BOOK{Heng:2017,
       author = {{Heng}, Kevin},
        title = "{Exoplanetary Atmospheres: Theoretical Concepts and Foundations}",
         year = 2017,
       adsurl = {https://ui.adsabs.harvard.edu/abs/2017eatc.book.....H}
}

@ARTICLE{Fortney:2005,
       author = {{Fortney}, Jonathan J.},
        title = "{The effect of condensates on the characterization of transiting planet atmospheres with transmission spectroscopy}",
      journal = {Monthly Notices of the Royal Astronomical Society},
         year = 2005,
        month = dec,
       volume = {364},
       number = {2},
        pages = {649-653},
          doi = {10.1111/j.1365-2966.2005.09587.x},
archivePrefix = {arXiv},
       eprint = {astro-ph/0509292},
 primaryClass = {astro-ph},
       adsurl = {https://ui.adsabs.harvard.edu/abs/2005MNRAS.364..649F}
}

@PHDTHESIS{MacDonald:2019,
       author = {{MacDonald}},
        title = "{Revealing the nature of exoplanetary atmospheres}",
       school = {University of Cambridge, UK},
         year = 2019,
        month = jan,
       adsurl = {https://ui.adsabs.harvard.edu/abs/2019PhDT.......177M}
}

@ARTICLE{Gardner:2023,
       author = {{Gardner}, Jonathan P. and {Mather}, John C. and {Abbott}, Randy and {Abell}, James S. and {Abernathy}, Mark and {Abney}, Faith E. and {Abraham}, John G. and {Abraham}, Roberto and {Abul-Huda}, Yasin M. and {Acton}, Scott and et al.},
        title = "{The James Webb Space Telescope Mission}",
      journal = {Publications of the Astronomical Society of the Pacific},
         year = 2023,
        month = jun,
       volume = {135},
       number = {1048},
          eid = {068001},
        pages = {068001},
          doi = {10.1088/1538-3873/acd1b5},
archivePrefix = {arXiv},
       eprint = {2304.04869},
 primaryClass = {astro-ph.IM},
       adsurl = {https://ui.adsabs.harvard.edu/abs/2023PASP..135f8001G}
}

@ARTICLE{Padovani:2023,
       author = {{Padovani}, Paolo and {Cirasuolo}, Michele},
        title = "{The Extremely Large Telescope}",
      journal = {Contemporary Physics},
         year = 2023,
        month = jan,
       volume = {64},
       number = {1},
        pages = {47-64},
          doi = {10.1080/00107514.2023.2266921},
archivePrefix = {arXiv},
       eprint = {2312.04299},
 primaryClass = {astro-ph.IM},
       adsurl = {https://ui.adsabs.harvard.edu/abs/2023ConPh..64...47P}
}

@ARTICLE{Skidmore:2015,
       author = {{Skidmore}, Warren and {TMT International Science Development Teams} and {Science Advisory Committee}, TMT},
        title = "{Thirty Meter Telescope Detailed Science Case: 2015}",
      journal = {Research in Astronomy and Astrophysics},
         year = 2015,
        month = dec,
       volume = {15},
       number = {12},
          eid = {1945},
        pages = {1945},
          doi = {10.1088/1674-4527/15/12/001},
archivePrefix = {arXiv},
       eprint = {1505.01195},
 primaryClass = {astro-ph.IM},
       adsurl = {https://ui.adsabs.harvard.edu/abs/2015RAA....15.1945S}
}

@INPROCEEDINGS{Fanson:2022,
       author = {{Fanson}, James and {Bernstein}, Rebecca and {Ashby}, David and {Bigelow}, Bruce and {Brossus}, Glenn and {Burgett}, William and {Demers}, Richard and {Fischer}, Barbara and {Figueroa}, Francisco and {Groark}, Frank and {Laskin}, Robert and {Millan-Gabet}, Rafael and {Park}, Samuel and {Pi}, Mart{\'\i} and {Turner}, Robert and {Walls}, Brian},
        title = "{Overview and status of the Giant Magellan Telescope project}",
    booktitle = {Ground-based and Airborne Telescopes IX},
         year = 2022,
       editor = {{Marshall}, Heather K. and {Spyromilio}, Jason and {Usuda}, Tomonori},
       series = {Society of Photo-Optical Instrumentation Engineers (SPIE) Conference Series},
       volume = {12182},
        month = aug,
          eid = {121821C},
        pages = {121821C},
          doi = {10.1117/12.2631694},
       adsurl = {https://ui.adsabs.harvard.edu/abs/2022SPIE12182E..1CF}
}

@ARTICLE{Snellen:2010,
       author = {{Snellen}, Ignas A.~G. and {de Kok}, Remco J. and {de Mooij}, Ernst J.~W. and {Albrecht}, Simon},
        title = "{The orbital motion, absolute mass and high-altitude winds of exoplanet HD209458b}",
      journal = {Nature},
         year = 2010,
        month = jun,
       volume = {465},
       number = {7301},
        pages = {1049-1051},
          doi = {10.1038/nature09111},
archivePrefix = {arXiv},
       eprint = {1006.4364},
 primaryClass = {astro-ph.EP},
       adsurl = {https://ui.adsabs.harvard.edu/abs/2010Natur.465.1049S}
}

@ARTICLE{Birkby:2017,
       author = {{Birkby}, J.~L. and {de Kok}, R.~J. and {Brogi}, M. and {Schwarz}, H. and {Snellen}, I.~A.~G.},
        title = "{Discovery of Water at High Spectral Resolution in the Atmosphere of 51 Peg b}",
      journal = {Astronomical Journal},
         year = 2017,
        month = mar,
       volume = {153},
       number = {3},
          eid = {138},
        pages = {138},
          doi = {10.3847/1538-3881/aa5c87},
archivePrefix = {arXiv},
       eprint = {1701.07257},
 primaryClass = {astro-ph.EP},
       adsurl = {https://ui.adsabs.harvard.edu/abs/2017AJ....153..138B}
}

@ARTICLE{Pelletier:2023,
       author = {{Pelletier}, Stefan and {Benneke}, Bj{\"o}rn and {Ali-Dib}, Mohamad and {Prinoth}, Bibiana and {Kasper}, David and {Seifahrt}, Andreas and {Bean}, Jacob L. and {Debras}, Florian and {Klein}, Baptiste and {Bazinet}, Luc and {Hoeijmakers}, H. Jens and {Kesseli}, Aurora Y. and {Lim}, Olivia and {Carmona}, Andres and {Pino}, Lorenzo and {Casasayas-Barris}, N{\'u}ria and {Hood}, Thea and {St{\"u}rmer}, Julian},
        title = "{Vanadium oxide and a sharp onset of cold-trapping on a giant exoplanet}",
      journal = {Nature},
         year = 2023,
        month = jul,
       volume = {619},
       number = {7970},
        pages = {491-494},
          doi = {10.1038/s41586-023-06134-0},
archivePrefix = {arXiv},
       eprint = {2306.08739},
 primaryClass = {astro-ph.EP},
       adsurl = {https://ui.adsabs.harvard.edu/abs/2023Natur.619..491P}
}

@ARTICLE{Burningham:2017,
       author = {{Burningham}, Ben and {Marley}, M.~S. and {Line}, M.~R. and {Lupu}, R. and {Visscher}, C. and {Morley}, C.~V. and {Saumon}, D. and {Freedman}, R.},
        title = "{Retrieval of atmospheric properties of cloudy L dwarfs}",
      journal = {Monthly Notices of the Royal Astronomical Society},
         year = 2017,
        month = sep,
       volume = {470},
       number = {1},
        pages = {1177-1197},
          doi = {10.1093/mnras/stx1246},
archivePrefix = {arXiv},
       eprint = {1701.01257},
 primaryClass = {astro-ph.SR},
       adsurl = {https://ui.adsabs.harvard.edu/abs/2017MNRAS.470.1177B}
}

@ARTICLE{Line:2013,
       author = {{Line}, Michael R. and {Wolf}, Aaron S. and {Zhang}, Xi and {Knutson}, Heather and {Kammer}, Joshua A. and {Ellison}, Elias and {Deroo}, Pieter and {Crisp}, Dave and {Yung}, Yuk L.},
        title = "{A Systematic Retrieval Analysis of Secondary Eclipse Spectra. I. A Comparison of Atmospheric Retrieval Techniques}",
      journal = {Astrophysical Journal},
         year = 2013,
        month = oct,
       volume = {775},
       number = {2},
          eid = {137},
        pages = {137},
          doi = {10.1088/0004-637X/775/2/137},
archivePrefix = {arXiv},
       eprint = {1304.5561},
 primaryClass = {astro-ph.EP},
       adsurl = {https://ui.adsabs.harvard.edu/abs/2013ApJ...775..137L}
}
\end{document}